\documentclass[12pt]{article}
\begin{document}
\newcommand{\beq}{\begin{equation}}
\newcommand{\eeq}{\end{equation}}
\newcommand{\beqa}{\begin{eqnarray}}
\newcommand{\eeqa}{\end{eqnarray}}
\newcommand{\beqar}{\begin{eqnarray*}}
\newcommand{\eeqar}{\end{eqnarray*}}
\newcommand{\al}{\alpha}
\newcommand{\be}{\beta}
\newcommand{\del}{\delta}
\newcommand{\D}{\Delta}
\newcommand{\eps}{\epsilon}
\newcommand{\ga}{\gamma}
\newcommand{\Ga}{\Gamma}
\newcommand{\ka}{\kappa}
\newcommand{\inn}{\!\cdot\!}
\newcommand{\h}{\eta}
\newcommand{\kk}{\varphi}
\newcommand\F{{}_3F_2}
\newcommand{\la}{\lambda}
\newcommand{\La}{\Lambda}
\newcommand{\na}{\nabla}
\newcommand{\Om}{\Omega}
\newcommand{\p}{\phi}
\newcommand{\sig}{\sigma}
\renewcommand{\t}{\theta}
\newcommand{\z}{\zeta}
\newcommand{\ssc}{\scriptscriptstyle}
\newcommand{\eg}{{\it e.g.,}\ }
\newcommand{\ie}{{\it i.e.,}\ }
\newcommand{\labell}[1]{\label{#1}} %{\label{#1}} %
\newcommand{\reef}[1]{(\ref{#1})}
\newcommand\prt{\partial}
\newcommand\veps{\varepsilon}
\newcommand\ls{\ell_s}
\newcommand\cF{{\cal F}}
\newcommand\cM{{\cal M}}
\newcommand\cN{{\cal N}}
\newcommand\cH{{\cal H}}
\newcommand\cC{{\cal C}}
\newcommand\cL{{\cal L}}
\newcommand\cG{{\cal G}}
\newcommand\cI{{\cal I}}
\newcommand\cl{{\iota}}
\newcommand\cP{{\cal P}}
\newcommand\cV{{\cal V}}
\newcommand\cg{{\it g}}
\newcommand\cR{{\cal R}}
\newcommand\cB{{\cal B}}
\newcommand\cO{{\cal O}}
\newcommand\tcO{{\tilde {{\cal O}}}}
\newcommand\bz{\bar{z}}
\newcommand\bw{\bar{w}}
\newcommand\hF{\hat{F}}
\newcommand\hA{\hat{A}}
\newcommand\hT{\hat{T}}
\newcommand\htau{\hat{\tau}}
\newcommand\hD{\hat{D}}
\newcommand\hf{\hat{f}}
\newcommand\hg{\hat{g}}
\newcommand\hp{\hat{\phi}}
\newcommand\hh{\hat{h}}
\newcommand\ha{\hat{a}}
\newcommand\hQ{\hat{Q}}
\newcommand\hP{\hat{\Phi}}
\newcommand\hb{\hat{b}}
\newcommand\hc{\hat{c}}
\newcommand\hd{\hat{d}}
\newcommand\hS{\hat{S}}
\newcommand\hX{\hat{X}}
\newcommand\tR{\tilde{R}}
\newcommand\tL{\tilde{\cal L}}
\newcommand\hL{\hat{\cal L}}
\newcommand\tG{{\widetilde G}}
\newcommand\tg{{\widetilde g}}
\newcommand\tphi{{\widetilde \phi}}
\newcommand\tPhi{{\widetilde \Phi}}
\newcommand\te{{\tilde e}}
\newcommand\tk{{\tilde k}}
\newcommand\tf{{\tilde f}}
\newcommand\tF{{\widetilde F}}
\newcommand\tK{{\widetilde K}}
\newcommand\tE{{\widetilde E}}
\newcommand\tpsi{{\tilde \psi}}
\newcommand\tX{{\widetilde X}}
\newcommand\tD{{\widetilde D}}
\newcommand\tO{{\widetilde O}}
\newcommand\tS{{\tilde S}}
\newcommand\tB{{\widetilde B}}
\newcommand\tA{{\widetilde A}}
\newcommand\tT{{\widetilde T}}
\newcommand\tC{{\widetilde C}}
\newcommand\tV{{\widetilde V}}
\newcommand\thF{{\widetilde {\hat {F}}}}
\newcommand\bR{{\textbf{R}}}
\newcommand\Tr{{\rm Tr}}
\newcommand\tr{{\rm tr}}
\newcommand\STr{{\rm STr}}
\newcommand\M[2]{M^{#1}{}_{#2}}
\parskip 0.3cm
%\begin{document}

%\thispagestyle{empty} \rightline{\small  \hfill IPM/P-2006/xxx}
\vspace*{1cm}

\begin{center}
{\bf \Large   Ramond-Ramond  corrections to type II supergravity at order $\alpha'^3$    }

\vspace*{1cm}

{  Mohammad R. Garousi\footnote{garousi@ferdowsi.um.ac.ir} }\\
\vspace*{1cm}
{ Department of Physics, Ferdowsi University of Mashhad,\\ P.O. Box 1436, Mashhad, Iran}
\\
\vspace{2cm}

\end{center}

\begin{abstract}
\baselineskip=18pt

Recently, it has been shown that  the  NS-NS corrections  to the type II supergravity   given by Gross and Sloan are invariant   under  the linear T-duality. In this paper, we study the invariance of this action under sequence of S-duality and  linear T-duality to find the R-R corrections to the supergravity at order $\alpha'^3$, up to field redefinition.

\end{abstract}
Keywords: T-duality, S-duality, effective action
%\end{document}
\setcounter{page}{0}
\setcounter{footnote}{0}
\newpage
%\beqa\frac\eeqa
\section{Introduction  } \label{intro}
It is known that the  type II superstring theory  is invariant under  T-duality \cite{Kikkawa:1984cp,TB,Giveon:1994fu,Alvarez:1994dn}  and  S-duality \cite{Font:1990gx,Sen:1994fa,Rey:1989xj,Sen:1994yi,Schwarz:1993cr,Hull:1994ys}.
The T-duality relates  type IIA superstring theory compactified on a circle with radius $\rho$ to  type IIB superstring theory compactified on another circle with radius $\alpha'/\rho$. It relates type IIA at weak (strong) coupling to type IIB at weak (strong) coupling. At low energy, this duality relates  type IIA supergravity  to  type IIB  supergravity. The S-duality, on the other hand, is a symmetry of type IIB superstring theory. It relates the type IIB theory at weak (strong) coupling to the type IIB at strong (weak) coupling. At low energy, this is the symmetry of type IIB  supergravity.   The higher derivative corrections to these supergravities  should have the same properties. That is, the higher derivatives of type IIA supergravity should be related to the higher derivatives of type IIB supergravity  under T-duality, and the higher derivatives of type IIB supergravity should be invariant under the S-duality.

The higher derivatives corrections to the supergravities start at order $\alpha'^3$ \cite{Gross:1986iv}. For  gravity, they involve couplings with structure $R^4$ where $R$ refers to the Riemann, the Ricci or the scalar curvatures. For B-field, they involve couplings with structure $(\prt H)^4$, $(\prt H)^2H^4$ or $H^8$. Similarly for dilaton and for mixed couplings. These higher derivative couplings appear in the contact terms of the corresponding string theory S-matrix elements at order $\alpha'^3$. For example, the couplings with structure $H^8$ appear in the contact terms of the S-matrix element of eight Kalb-Ramond vertex operators. The evaluation of this eight-point function and the extraction of  its contact terms, however,  are  extremely difficult tasks. So one has to take advantage of  the symmetries/dualities  of the effective field theory to find such higher derivative couplings. Supersymmetry is able to determine all  $\alpha'^3$-couplings   up to field redefinitions \cite{Howe:1983sra,Nilsson:1986rh,Paban:1998ea,Green:1998by}. It can also fix the moduli-dependence of the type IIB theory \cite{Green:1998by}. In this paper, we are going to demonstrate that the T-duality and S-duality  are also able to find all  $\alpha'^3$-couplings and  fix the moduli-dependence of the type IIB theory.

   The Riemann curvature couplings  at order $\alpha'^3$ have been found in \cite{ Gross:1986iv} by analyzing the sphere-level four-graviton scattering amplitude in type II superstring theory. The result in the eight-dimensional transverse space of the light-cone formalism, is a polynomial in the Riemann curvature tensors 
\beqa
L&\sim& t^{i_1\cdots i_8}t_{j_1\cdots j_8}R_{i_1i_2}{}^{j_1j_2}\cdots R_{i_7i_8}{}^{j_7j_8}+\cdots \labell{Y1}
\eeqa
where $t^{\{\cdots\}}$ is a tensor in eight dimensions which includes the eight-dimensional Levi-Civita tensor \cite{Gross:1986iv} and dots represent terms containing the Ricci and scalar curvature tensors. They can not be captured by the four-graviton scattering amplitude as they are zero on-shell. 
There are two expressions in the literature for the  Lorentz invariant extension of the  above $SO(8)$ invariant  Lagrangian. One of them which is consistent with the sigma model beta function is   the following   \cite{Grisaru:1986vi,Freeman:1986zh}:
\beqa
\cL'&=&   R_{hmnk}R_p{}^{mn}{}_qR^{hrsp}R^q{}_{rs}{}^k+\frac{1}{2}R_{hkmn}R_{pq}{}^{mn}R^{hrsp}R^q{}_{rs}{}^k+\cdots\labell{Y2}
\eeqa
where   dots represent the specific form of the  off-shell  Ricci and scalar curvature couplings which reproduce the sigma model beta function \cite{Grisaru:1986vi,Freeman:1986zh}.  The sigma-model   requires the Levi-Civita tensors in $t_8t_8$ to appear in the action   as the four Riemann curvature couplings $\eps_{10}\cdot\eps_{10}RRRR$   \cite{Grisaru:1986vi,Freeman:1986zh}. This term has its first non-zero contribution at  five gravitons \cite{Zumino:1985dp}. It has been shown in  \cite{Garousi:2012yr} that the above Lagrangian is not consistent with the standard form of the T-duality transformations. 

The other Lorentz invariant extension for the  $SO(8)$ invariant Lagrangian \reef{Y1} has been given in \cite{Gross:1986mw}:
\beqa
\cL(R)&=&R_{hkmn}R^{krnp}R_{rs}{}^{qm}R^{sh}{}_{pq}+\frac{1}{2}R_{hkmn}R^{krnp}R_{rspq}R^{shqm}\labell{Y3}\\
&&  -\frac{1}{2}R_{hkmn}R^{krmn}R_{rspq}R^{shpq} -\frac{1}{4}R_{hkmn}R^{krpq}R_{rs}{}^{mn}R^{sh}{}_{pq}\nonumber\\
&& +\frac{1}{16}R_{hkmn}R^{khpq}R^{rsmn}R_{srpq}+\frac{1}{32}R_{hkmn}R^{khmn}R_{rspq}R^{srpq}\nonumber
\eeqa
 It has been shown in \cite{Myers:1987qx} that, up to field redefinition, the difference between the above two Lagrangians is the   couplings $\eps_{10}\cdot\eps_{10}RRRR$. The B-field and dilaton couplings have been added to the above Lagrangian by extending the Riemann curvature to the generalized Riemann curvature \cite{Gross:1986mw}. The resulting  Lagrangian is then invariant under standard linear T-duality  transformations \cite{Garousi:2012jp}. The above sphere-level gravity couplings can be extended to the S-duality invariant form by including the loops and nonperturbative corrections  \cite{Green:1997tv} - \cite{Basu:2007ck}.
 
The  generalized Riemann curvature  does not include the Ramond-Ramond (R-R) fields, so the above Lagrangian for the generalized  Riemann curvature is not consistent with S-duality even at the sphere-level. In this paper, we are going to finds the couplings for the R-R fields by requiring the action to be invariant under both S-duality and T-duality. Consistency of D-brane effective action with T-duality and S-duality has been used in   \cite{Garousi:2009dj,Garousi:2010ki,Becker:2010ij,Garousi:2010rn,Garousi:2010bm,Velni:2012sv,McOrist:2012yc,Kahle:2013kg} and    \cite{Bachas:1999um,Garousi:2011fc,Garousi:2012gh}     to find new couplings in the D-brane effective actions.

The outline of the paper is as follows: We begin in section 2 by reviewing the linear T-duality transformations.   In section 3, we review the   S-duality transformations. In section 4, we find the couplings of two Riemann curvatures and two R-R three-form field strengths by requiring the couplings with structure $HHRR$ in \reef{Y3} to be extended to the $SL(2,R)$ invariant form. Then by applying T-duality on them we  find couplings with structure $F^{(n)}F^{(n)}RR$ for $n=1,2,3,4,5$ . In section 5, we first  find the couplings between two R-R three-form field strengths and two B-field strengths by requiring their consistency with the couplings found in section 4. In the course of this study we  have  used on-shell relations. Then by applying T-duality on them we find the couplings with structure $F^{(n)}F^{(n)}HH$. In section 6, we find couplings with structure  $F^{(n)}F^{(n+2)}HR$. In section 7, we apply T-duality on the couplings found in section 6 to find the couplings with structure $F^{(n)}F^{(n+4)}RR$. We find that these couplings are zero, as expected. In section 8, we apply T-duality on the couplings found in section 6 to find the couplings with structure $F^{(n)}F^{(n+4)}HH$. In all sections, we make the couplings in the type IIB theory to be invariant under the S-duality. This permits us   to find the  couplings of four R-R field strengths, as well. In section 9, we briefly discuss our results.

\section{T-duality}

The full set of T-duality transformations for massless R-R and NS-NS fields in type II superstring theories has been found in \cite{TB,Meessen:1998qm,Bergshoeff:1995as,Bergshoeff:1996ui,Hassan:1999bv}.  The nonlinear T-duality transformations of the  fields $C$ and $B$ are such that  the expression ${\cal C}=e^BC$ has  a linear transformation under the T-duality \cite{Taylor:1999pr}. When  the T-duality transformation acts along the Killing coordinate $y$,  these   fields   transform as:
\beqa
e^{2\tPhi}=\frac{e^{2\Phi}}{G_{yy}}&;& 
\tG_{yy}=\frac{1}{G_{yy}}\nonumber\\
\tG_{\mu y}=\frac{B_{\mu y}}{G_{yy}}&;&
\tG_{\mu\nu}=G_{\mu\nu}-\frac{G_{\mu y}G_{\nu y}-B_{\mu y}B_{\nu y}}{G_{yy}}\nonumber\\
\tB_{\mu y}=\frac{G_{\mu y}}{G_{yy}}&;&
\tB_{\mu\nu}=B_{\mu\nu}-\frac{B_{\mu y}G_{\nu y}-G_{\mu y}B_{\nu y}}{G_{yy}}\labell{nonlinear}\\
{\cal \tC}^{(n)}_{\mu\cdots \nu y}={\cal C}^{(n-1)}_{\mu\cdots \nu }&;&
{\cal \tC}^{(n)}_{\mu\cdots\nu}={\cal C}^{(n+1)}_{\mu\cdots\nu y}\nonumber
\eeqa
where $\mu,\nu$ denote any coordinate directions other than $y$. In above transformation the metric is given in the string frame. If $y$ is identified on a circle of radius $\rho$, \ie $y\sim y+2\pi \rho$, then after T-duality the radius becomes $\tilde{\rho}=\alpha'/\rho$. The string coupling $g=e^{\phi_0}$ is also shifted as $\tilde{g}=g\sqrt{\alpha'}/\rho$.

We would like to study the T-duality of the $\alpha'^3$-order couplings of type II superstring theory which involve four massless fields, so we need the above transformations at the linear order. Assuming that the  massless  fields are small perturbations around the background, \ie
\beqa
G_{\mu\nu}&=&\eta_{\mu\nu}+2\kappa h_{\mu\nu}\,;\,\,G_{yy}\,=\,\frac{\rho^2}{\alpha'}(1+2\kappa h_{yy})\,;\,\,G_{\mu y}\,=\,2\kappa h_{\mu y}\nonumber\\
B^{(2)}&=&2\kappa b^{(2)}\,;\,\,
\Phi\,=\,\phi_0+\sqrt{2}\kappa \phi\,;\,\,{\cal \tC}^{(n)}\,=\,2\kappa c^{(n)}\labell{perturb}
\eeqa
  the   transformations \reef{nonlinear} take the following linear form for the perturbations:
\beqa
&&
 \sqrt{2}\tilde{\phi}=\sqrt{2}\phi - h_{yy},\,\tilde{h}_{yy}=-h_{yy},\, \tilde{h}_{\mu y}=b_{\mu y}/\eta_{yy},\, \tilde{b}_{\mu y}=h_{\mu y}/\eta_{yy},\,\tilde{h}_{\mu\nu}=h_{\mu\nu},\,\tilde{b}_{\mu\nu}=b_{\mu\nu}\nonumber\\
&&{\tilde{C}}^{(n)}_{\mu\cdots \nu y}={ C}^{(n-1)}_{\mu\cdots \nu },\,\,\,{ \tilde{C}}^{(n)}_{\mu\cdots\nu}={ C}^{(n+1)}_{\mu\cdots\nu y}\labell{linear}
\eeqa
  
The effective action at order $\alpha'^3$ includes the Riemann curvature,  the derivative of  the B-field strength $H=\frac{1}{2}dB$ and the derivative of the R-R field strength $F=\frac{1}{2}dC$ . So we need the transformation of these fields under the linear T-duality.  The transformation of the linearized Riemann curvature tensor  when it carries zero, one and two Killing indices are, respectively, \cite{Garousi:2012yr}\footnote{The metric $\eta_{yy}=\rho^2/\alpha'$. In \cite{Garousi:2012yr}, the T-duality has been considered for the self-dual radius, \ie $\eta_{yy}=1$.}  
\beqa
R_{abcd}&\leftrightarrow & R_{abcd}\nonumber\\
R_{abc y}\rightarrow  -  H_{ab y,c}/\eta_{yy}&;&H_{ab y,c}\rightarrow  -R_{abc y}/\eta_{yy}\labell{TR}\\
R_{ayby}&\rightarrow &-R_{ayby}/(\eta_{yy})^2\nonumber
\eeqa
where the linearized Riemann curvature and $H$ are
\beqa
R_{abcd}&=& \kappa(h_{ad,bc}+h_{bc,ad}-h_{ac,bd}-h_{bd,ac})\nonumber\\
H_{abc}&=&\kappa(b_{ab,c}+b_{ca,b}+b_{bc,a})\labell{RH}
\eeqa
where as usual  the comma represents partial differentiation. The field strength $H$ is also  invariant when it carries no Killing index.  The T-duality of the R-R field strength  can easily be read from the transformations \reef{linear} to be 
\beqa
F^{(n)}_{\mu\cdots\nu}\leftrightarrow F^{(n+1)}_{\mu\cdots\nu y}\labell{TF}
\eeqa
where we have used the fact that the fields are independent of the $y$-direction.

%As we will see in the next section the effective action in the string frame involves the second derivative of the dilaton which is not invariant under the T-duality. It must be combined with the Ricci curvature to become invariant \cite{Garousi:2012yr}. The transformation of the Ricci curvature when it carries zero, one and two $y$ indices are, respectively, \cite{Garousi:2012yr} 
%\beqa
%R_{ab}+2\sqrt{2}\kappa \phi_{,ab}&\leftrightarrow &R_{ab}+2\sqrt{2}\kappa \phi_{,ab}\labell{TR2}\\
% R_{ay}&\leftrightarrow &-H_{ay} \nonumber\\
%R_{yy}&\leftrightarrow& -R_{yy}\nonumber 
%\eeqa
%where $H_{ab}$ is the divergence of the B-field strength, \ie $H_{ab} \equiv H_{abc}{}^{;c}$ where the semicolon represents covariant differentiation.

In a T-duality invariant theory, all the couplings in the dimensional reduction must be invariant under the above T-duality transformations. So to extend a coupling to a set of couplings which   
  are invariant under linear T-duality,    we first use the dimensional reduction to reduce the 10-dimensional couplings to 9-dimensional couplings, \ie separate the  indices  along and orthogonal to $y$,     and then apply the  above T-duality transformations.  If the original coupling is not invariant under the T-duality, one must add new terms to make them invariant.  
  
%Before ending this section, let us compare the above  method for finding the T-duality invariant couplings in the  spacetime, and the method used in \cite{Garousi:2009dj} for finding the T-duality invariant couplings on the D-brane world volume. In both cases, one first uses  the dimensional reduction to separate the indices along and orthogonal to the Killing $y$-direction, then uses the linear T-duality transformations. After the T-duality has been performed, the indices must be complete.  In a T-duality invariant theory in spacetime, the completion of the $y$-index trivially gives the original theory because the $y$-index is a spacetime index before and after T-duality. However, in a D-brane theory the world volume Killing index $y$ becomes the transverse index after T-duality. Hence, the completion of the transverse indicies gives a nontrivial constraint on the D-brane couplings \cite{Garousi:2009dj}.  

\section{S-duality}

Unlike the T-duality transformations \reef{nonlinear} which are given in string frame, the S-duality transformations of type IIB theory are properly defined in the Einstein frame , $G^s_{ab}=G_{ab}e^{\Phi/2}$. In this frame  the metric and the R-R four-form are invariant. The B-field and the R-R two-form transform as doublet \cite{ Tseytlin:1996it,Green:1996qg}. Since the parameters of the duality are constant, their field strengths, \ie $H=\frac{1}{2}dB$ and $F=\frac{1}{2}dC$,  are  also transformed as doublet, 
\beqa
\cH\equiv\pmatrix{H \cr 
F}\rightarrow (\Lambda^{-1})^T \pmatrix{H \cr 
F}\,\,\,;\,\,\,\Lambda=\pmatrix{p&q \cr 
r&s}\in SL(2,R)\labell{2}
\eeqa
The dilaton and the R-R scalar transform nonlinearly as $\tau\rightarrow \frac{p\tau+q}{r\tau+s}$ where the complex scalar field is defined as $\tau=C+ie^{-\Phi}$. The matrix $\cM$ defined in terms of the dilaton and the R-R scalar, \ie 
 \beqa
 {\cal M}=e^{\Phi}\pmatrix{|\tau|^2&C \cr 
C&1}\labell{M}
\eeqa
then  transforms  as \cite{Gibbons:1995ap}
\beqa
{\cal M}\rightarrow \Lambda {\cal M}\Lambda ^T
\eeqa
The derivatives of this matrix also transform as above. The matrix $\cN$ which is defined as 
 \beqa
 {\cal N}=\pmatrix{0&1 \cr 
-1&0}\labell{N}
\eeqa
  can be written as $
\Lambda {\cal N}\Lambda ^T$. Using these matrices and the transformation \reef{2}, one can construct different $SL(2,R)$ invariant objects.

An $SL(2,R)$ invariant object, in general, has more than one  components. For example,  $\cH^T\cM_0\cH$ has the following components: 
\beqa
 \cH^T\cM_0\cH =e^{-\phi_0}(1+e^{2\phi_0}C_0^2)HH+e^{\phi_0}FF-e^{\phi_0}C_0(HF+FH)\labell{SHH}
\eeqa
where the index zero in various fields refers to the constant background value of these fields. If one finds somehow one of the above tree-level components, then the requirement that it should be invariant under the S-duality gives all other tree-level components automatically. 

 In this paper we are going to find an action which is invariant under  both T-duality and S-duality. One of the dualities  is in the Einstein frame in which the dilaton appears and the other one is in the string frame in which the  dilaton appears only through  an overall factor. So the  question is how to transform an S-duality invariant coupling, \eg \reef{SHH}, under T-duality? For the components which do not include the background R-R scalar, one has to simply move back to the string frame and apply the T-duality on them.  For the components in which the nonzero $C_0$ appears, one has to transform them to zero coupling because the type IIB background with non-zero $C_0$  transforms to the type IIA with background $C^{y}_0$. However, the constant vector is gauged away by the the R-R gauge symmetry. Therefore, to apply T-duality on an S-duality invariant coupling, one should first set the background R-R scalar to zero and then move to the string frame to apply the standard T-duality transformations on it.
 
The S-duality of  the Riemann curvature corrections to the supergravity has been studied in \cite{Green:1997tv} - \cite{Basu:2007ck}. The Riemann curvature in the Einstein frame is invariant under the $SL(2,R)$ transformations, and the overall dilaton factor which is not invariant under the $SL(2,R)$ transformation has been extended to the non-holomorphic Eisenstein series $E_{3/2}(\tau_0,\bar{\tau}_0)$ which is invariant under the $SL(2,Z)$ transformations. In his paper, we will see that this structure appear for all other four-field couplings. That is, the couplings can be written as two factors. One factor is the series  $E_{3/2}(\tau_0,\bar{\tau}_0)$ which has the effect of loops and nonperturbative effects, and the other factor is the tree-level couplings which are invariant under the $SL(2,R)$ transformations.

\section{$F^{(n)}F^{(n)}RR$ }

The couplings of two Riemann curvatures and two identical R-R field strengths can be found by requiring the existing couplings in the literature to be consistent with S-duality and T-duality.  The couplings  of four NS-NS states in the Einstein frame  have been found in \cite{Gross:1986mw,Myers:1987qx}
 \beqa
 S= \frac{\gamma}{\kappa^2}\int d^{10}x\sqrt{-G} e^{-3\phi_0/2}\cL(\bar{R})\labell{S}
 \eeqa
 where $\gamma=\frac{\alpha'^3}{2^5} \z(3)$, the   Lagrangian $\cL $ is given in \reef{Y3}
% \beqa
%\cL(\bar{R})&=&  \big[\bar{R}_{hkmn}\bar{R}_{krnp}\bar{R}_{rs}{}_{qm}\bar{R}_{sh}{}_{pq}+\frac{1}{2}\bar{R}_{hkmn}\bar{R}_{krnp}\bar{R}_{rspq}\bar{R}_{shqm}\labell{Y3}\\
%&& -\frac{1}{2}\bar{R}_{hkmn}\bar{R}_{krmn}\bar{R}_{rspq}\bar{R}_{shpq} -\frac{1}{4}\bar{R}_{hkmn}\bar{R}_{krpq}\bar{R}_{rs}{}_{mn}\bar{R}_{sh}{}_{pq}\nonumber\\
%&&+\frac{1}{16}\bar{R}_{hkmn}\bar{R}_{khpq}\bar{R}_{rsmn}\bar{R}_{srpq}+\frac{1}{32}\bar{R}_{hkmn}\bar{R}_{khmn}\bar{R}_{rspq}\bar{R}_{srpq}\bigg]\nonumber
%\eeqa 
and the generalized Riemann curvature is \cite{Gross:1986mw}, 
\beqa
 \bar{R}_{ab}{}^{cd}&= &R_{ab}{}^{cd}-\frac{\kappa}{\sqrt{8}}\eta_{[a}{}^{[c}\phi_{;b]}{}^{d]}+  e^{-\phi_0/2}H_{ab}{}^{[c;d]}\labell{trans}
\eeqa
where   the bracket notation is $H_{ab}{}^{[c;d]}=H_{ab}{}^{c;d}-H_{ab}{}^{d;c}$. 
In the string frame the   linearized $\bar{R}_{ab cd}$  becomes \cite{Garousi:2012jp}
\beqa
\bar{R}_{ab cd}=e^{-\phi_0/2}\cR_{ab cd}
\eeqa
where $\cR_{ab cd}$ is the following expression 
\beqa
\cR_{ab cd}&=&R_{ab cd}+H_{ab [c,d]}\labell{RH2}
\eeqa
The dilaton appears only as the overall   factor  $e^{-2 \phi_0}\sqrt{-G}$  which is invariant under nonlinear T-duality. It has been shown in \cite{Garousi:2012jp} that the  Lagrangian $\cL(\cR)$ is  invariant under   linear T-duality transformation \reef{TR}.
The action \reef{S} does not include the R-R fields so obviously it is not invariant under the S-duality. I this paper, we are  going to find new couplings involving the R-R fields to  make the above action  invariant under both S-duality and T-duality.

The Lagrangian $\cL$ has couplings with structures $R^4$, $H^4$ and $R^2H^2$ in the string frame. However, in the Einstein frame which is the appropriate frame for studying S-duality, this Lagrangian includes the couplings with structure $R^4$, $ R^2\phi^2$, $\phi^4$, $ H^4$, $H^2\phi^2$, $H^2\phi R$ and $R^2H^2$. The $R^4$ couplings are invariant under the S-duality, however, all other terms must be combined with appropriate R-R couplings to be invariant under the $SL(2,R)$ transformations. We begin by extending the couplings $R^2H^2$ to S-duality invariant form to find the couplings of two Riemann curvatures and two R-R three-form field strengths in type IIB theory. 

The couplings of two Riemann curvatures and two B-field strengths  can  be extended to the $SL(2,R)$ invariant form by using the following replacement:
\beqa
e^{-\phi_0}HHR^2\rightarrow \cH^T\cM_0\cH R^2
\eeqa
which produces new tree-level couplings. The overall dilaton factor in \reef{S} is  extended to the $SL(2,Z)$ invariant non-holomorphic Eisenstein series $E_{3/2}(\tau_0,\bar{\tau}_0)$ which produces  the loops and nonperturbative effects \cite{Green:1997tv} - \cite{Basu:2007ck}. To apply T-duality on the resulting S-duality invariant couplings, one should set the background R-R scalar to zero and move back to the string frame. One finds the following tree-level couplings between two Riemann curvatures and two R-R three-form field strengths\footnote{From now on, we use only subscript indices and the repeated indices are contracted with the   metric.}:
\beqa
S &\supset&\frac{\gamma}{\kappa^2} \int d^{10}x\sqrt{-G}\bigg[\frac{1}{16} F_{r s[p,q]}^2 R_{h k m n}^2+\frac{1}{4} F_{r s[m,n]} F_{r s[p,q]} R_{h k m n} R_{h k p q}\nonumber\\&&+F_{h s[p,q]} F_{r s[p,q]} R_{h k m n} R_{k r m n}-2 F_{h s[q,m]} F_{r s[p,q]} R_{h k m n} R_{k r n p}-2 F_{h s[p,q]} F_{r s[q,m]} R_{h k m n} R_{k r n p}\nonumber\\&&+F_{h s[p,q]} F_{r s[m,n]} R_{h k m n} R_{k r p q}+F_{k r[m,n]} F_{r s[p,q]} R_{h k m n} R_{h s p q}+\frac{1}{2} F_{k r[p,q]} F_{h s[p,q]} R_{h k m n} R_{r s m n}\nonumber\\&&+\frac{1}{8} F_{h k[m,n]} F_{r s[p,q]} R_{h k p q} R_{r s m n}+4 F_{k r[n,p]} F_{h s[p,q]} R_{h k m n} R_{r s m q}+F_{k r[m,n]} F_{h s[p,q]} R_{h k m n} R_{r s p q}\nonumber\\&&-F_{k r[n,p]} F_{h s[q,m]} R_{h k m n} R_{r s p q}\bigg]\labell{F3R}
\eeqa
From one loop string amplitude, a representation for the  $R^2(\prt F^{(3)})^2$ couplings has been found in \cite{Peeters:2003pv} which reproduces the  above couplings, up to field redefinition. The field redefinition means the couplings involving the Ricci and scalar curvatures, as well as the divergence of the form-field strengths  can be absorbed into supergravity by field redefinition involving these higher derivative terms \cite{Gross:1986iv}.

We now apply the  T-duality transformations on the above couplings to find the  new couplings for two R-R two-form field strengths and two Riemann curvatures in type IIA theory. To this end,  we use the dimensional reduction on the above couplings and  find the couplings with structure $\eta^{yy}F^{(3)}_yF^{(3)}_yRR$ where $F^{(3)}_y$ is the R-R field strength which carries one Killing index and $R$ is the Riemann curvature which carries no Killing index. Under the T-duality transformations \reef{TR} and \reef{TF} this structure transforms to $\eta_{yy}F^{(2)}F^{(2)}RR$. The square root of the determinant of the metric $\sqrt{-G}$ also transforms as $\sqrt{-G}\eta^{yy}$. So one finds the couplings which have no Killing index. The integral over the circle which gives $2\pi\rho$ transforms to $2\pi\alpha'/\rho$. The    ten-dimensional Newton constant $\kappa^2$ also transforms to $\kappa^2\alpha'/\rho^2$.  The result for all terms in \reef{F3R} are   the following:
\beqa
S &\supset&\frac{\gamma}{\kappa^2} \int d^{10}x\sqrt{-G}\bigg[ \frac{1}{4} F_{p q,r}^2 R_{h k m n}^2+\frac{1}{2} F_{m n,s} F_{p q,s} R_{h k m n} R_{h k p q}+F_{p q,h} F_{p q,r} R_{h k m n} R_{m n k r}\nonumber\\&&+2 F_{h s,q} F_{r s,q} R_{h k m n} R_{m n k r}+F_{k r,q} F_{h s,q} R_{h k m n} R_{m n r s}+2 F_{m q,r} F_{p q,h} R_{h k m n} R_{n p k r}\nonumber\\&&+2 F_{m q,h} F_{p q,r} R_{h k m n} R_{n p k r}+2 F_{h s,p} F_{r s,m} R_{h k m n} R_{n p k r}+2 F_{h s,m} F_{r s,p} R_{h k m n} R_{n p k r}\nonumber\\&&-4 F_{k r,n} F_{h s,q} R_{h k m n} R_{m q r s}+F_{m n,r} F_{p q,h} R_{h k m n} R_{p q k r}-F_{m n,k} F_{p q,s} R_{h k m n} R_{p q h s}\bigg]\labell{F2R}
\eeqa
The indices run over all directions except the $y$-direction and the fields are all independent of the $y$-direction. One of the integrals is also over the  original  circle with radius $ \rho$. After using the T-duality, one  assumes the fields depend on all directions and the indices run over all spacetime directions. Hence, the above  couplings are the couplings of two Riemann curvatures and two R-R two-form field strengths that are predicted by T-duality. 

It has been observed in \cite{Garousi:2012yr,Barreiro:2012aw} that the S-matrix element of four NS-NS vertex operators at order $\alpha'^3$ does not produce term which has zero or one Mandelstam variable $k_i\inn k_j$. Using the S-dual and T-dual Ward identity of S-matrix elements \cite{Garousi:2011we}, one expects all four-point functions share this property. Using the above action, we have calculated the S-matrix element of two gravitons and two R-R one-forms and observed that the amplitude does not produce  term with zero or one factor of the Mandelstam variable $k_i\inn k_j$.

Using the above ten-dimensional couplings one can perform again the  dimensional reduction on a circle and finds  the couplings with structure $\eta^{yy}F^{(2)}_yF^{(2)}_yRR$. Doing the same steps as above, one finds the following couplings between two Riemann curvatures and two R-R one-form field strengths in the  string frame:
\beqa
S &\supset&\frac{\gamma}{\kappa^2} \int d^{10}x\sqrt{-G}\bigg[\frac{1}{2} F_{q,r}^2 R_{h k m n}^2+4 F_{p,h} F_{p,r} R_{h k m n} R_{m n k r}\nonumber\\&&+4 F_{m,r} F_{p,h} R_{h k m n} R_{n p k r}+4 F_{m,h} F_{p,r} R_{h k m n} R_{n p k r}\bigg]\labell{F1R}
\eeqa
where we have used the fact that the two indices in the R-R field strength are symmetric, \ie $F_{q,r}=\frac{1}{2}C_{,qr}$. %While we expect the couplings \reef{F2R} to be reproduced fully by the appropriate string theory S-matrix element, we do not expect the S-matrix element produce correctly the overall coefficient of the above couplings. The reason is that  the S-duality extension of the Lagrangian $\cL_2$ in \reef{S} produces  similar terms as above.

The above couplings are in type IIB theory, so one can extend them to the S-duality invariant forms.  
  In the Einstein frame, one finds   the overall dilaton factor to be $e^{\phi_0/2}$. This can be written as $e^{-3\phi_0/2}\times e^{2\phi_0}$. The first factor is the  dilaton factor which appears also in $R^4$ terms. This factor should be extended to the  non-holomorphic Eisenstein series $E_{3/2}(\tau_0,\bar{\tau}_0)$ after including the loops and nonperturbative effects. The second factor must be combined with the R-R fields to be extended to a $SL(2,R)$ invariant form. Using the matrix $\cM$, one finds the following $SL(2,R)$ invariant
\beqa
-\frac{1}{4}\Tr[\cM_{,hk}\cM^{-1}_{,mn} ]&=&(2e^{2\phi_0}F_{h,k}F_{m,n}+\kappa^2\phi_{,hk}\phi_{,mn})\labell{FF}
\eeqa
Therefore, the 
S-duality invariant action  is the following:
 \beqa
S &\supset&\frac{\gamma}{\kappa^2} \int d^{10}x\sqrt{-G}E_{3/2}\bigg[-\frac{1}{16}\Tr[\cM_{,qr}\cM^{-1}_{,qr} ] R^2_{h k mn}-\frac{1}{2} \Tr[\cM_{,ph}\cM^{-1}_{,pr} ] R_{h k m n}R_{mnkr} \nonumber\\&&-\frac{1}{2}  \Tr[\cM_{,mr}\cM^{-1}_{,ph} ] R_{h k m n}R_{npkr} -\frac{1}{2}  \Tr[\cM_{,mh}\cM^{-1}_{,pr} ] R_{h k m n}R_{npkr}\bigg]\labell{F1R2}
\eeqa
 One can easily observe that the above action  reproduces  the couplings \reef{F1R}. This action also produces the tree-level couplings  between  two Riemann curvatures and two dilatons
\beqa
S &\supset&\frac{\gamma}{\kappa^2} \int d^{10}x\sqrt{-G}e^{-3\phi_0/2}\kappa^2\bigg[\frac{1}{4} \phi_{,qr}^2 R_{h k m n}^2+2 \phi_{,ph} \phi_{,pr} R_{h k m n} R_{m n k r}\nonumber\\&&+2 \phi_{,mr} \phi_{,ph} R_{h k m n} R_{n p k r}+2 \phi_{,mh} \phi_{,pr} R_{h k m n} R_{n p k r}\bigg]\labell{F1R1}
\eeqa
We have checked that the above couplings are  exactly the $\phi^2R^2$ couplings in  the action \reef{S}. Therefore, the combination of the $\phi^2R^2$ couplings in \reef{S} and the couplings \reef{F1R} is invariant under the $SL(2,R)$ transformations. There would be no new couplings by applying T-duality on \reef{F1R2}.

There have been  no ambiguity in finding the couplings \reef{F2R} and \reef{F1R} from the dimensional reduction of \reef{F3R} because the $y$-index does not appear in the T-dual couplings. However, in finding the couplings $F^{(4)}F^{(4)}RR$ from \reef{F3R}, the $y$-index appear in the T-dual couplings which makes the final couplings to be ambitious. In this case we need other information to find the couplings correctly.  
 We will use the observation that the four-point functions has no term with zero and one Mandelstam variable \cite{Garousi:2012yr,Barreiro:2012aw},  to fix the above ambiguity.

To find the couplings $F^{(4)}F^{(4)}RR$ in type IIA theory from the type IIB couplings \reef{F3R}, we use  the dimensional reduction on the ten-dimensional action \reef{F3R} and keep the terms which have no $y$-index. Then under the T-duality \reef{TF} the R-R three-form field strength transforms to the four-form field strength with one $y$-index. As a result, one finds the couplings with structure $\eta^{yy}F^{(4)}_yF^{(4)}_yRR$ where the metric $\eta^{yy}$ is coming from the T-duality of $\sqrt{-G}$. Then one can easily find the $F^{(4)}F^{(4)}RR$ couplings in which the R-R field strengths contract at least once. This can be done  by converting the $y$-index to complete spacetime index and taking into account the appropriate symmetry factor. To clarify the symmetry factor consider,   for example, a coupling which has three contracted indices between the R-R field strengths. The  symmetry factor is $1/3$ for that term. The result for all couplings in \reef{F3R} is
\beqa
S &\supset&\frac{\gamma}{\kappa^2} \int d^{10}x\sqrt{-G} \bigg[\frac{1}{32} F_{p r s t,q}^2 R_{h k m n}-\frac{1}{24} F_{p r s t,q} F_{q r s t,p} R_{h k m n}+\frac{1}{3} F_{m r s t,n} F_{p r s t,q} R_{h k p q}\nonumber\\&&-\frac{2}{3} F_{h p s t,q} F_{p r s t,q} R_{m n k r}+F_{h p s t,q} F_{q r s t,p} R_{m n k r}+\frac{1}{2} \left(F_{k p r t,q}-2 F_{k q r t,p}\right) F_{h p s t,q} R_{m n r s}\nonumber\\&&+\left(F_{h p s t,q}-F_{h q s t,p}\right) F_{n r s t,q} R_{m p k r}+\left(F_{h n s t,q}-F_{h q s t,n}\right) F_{p r s t,q} R_{m p k r}\nonumber\\&&-\left(F_{h p s t,q}-\frac{2}{3} F_{h q s t,p}\right) F_{q r s t,n} R_{m p k r}-\left(F_{h n s t,q}-\frac{2}{3} F_{h q s t,n}\right) F_{q r s t,p} R_{m p k r}\nonumber\\&&-2 \left(-2 F_{k n r t,p}+F_{k p r t,n}\right) F_{h p s t,q} R_{m q r s}-4 \left(F_{k n r t,p}-F_{k p r t,n}\right) F_{h q s t,p} R_{m q r s}\nonumber\\&&-2 F_{h q s t,p} F_{n r s t,m} R_{p q k r}-2 F_{k n r t,m} F_{q r s t,p} R_{p q h s}+F_{k n r t,p} F_{h m s t,q} R_{p q r s}\nonumber\\&&+\left(-2 F_{k n r t,p}+F_{k p r t,n}\right) F_{h q s t,m} R_{p q r s}+4 F_{k n r t,m} F_{h q s t,p} R_{p q r s}\nonumber\\&&+\frac{1}{2} F_{h k q t,p} F_{n r s t,m} R_{p q r s}-\frac{3}{4} F_{h k p q,s} F_{m n r t,s} R_{p q r t}\bigg]R_{h k m n}\labell{F4R}
\eeqa
All terms except the last one are the  couplings in which the R-R field strengths contract with each other. However, they are not the whole couplings between two R-R four-form field strengths and two Riemann curvatures. The  $F^{(4)}F^{(4)}RR$ couplings which have no contraction between the two R-R four-forms do not produce the couplings $F^{(4)}_yF^{(4)}_yRR$ under the dimensional reduction. To find such couplings we  should first use the Bianchi identity to find the independent ones. One observes that there are two independent couplings which have no contraction between the two R-R four-forms. One of them is the coupling in the last term of \reef{F4R} and the other one is $ F_{trsp,q} F_{mnhq,p} R_{trhk}R_{mnsk}$. 

To implement the Bianchi identities, one may  write  the R-R field strengths and the linearized Riemann curvatures \reef{RH} in terms of the R-R potential and the metric fluctuation $h_{\mu\nu}$, respectively. The symmetry of the metric fluctuation can be implemented by writing it as $h_{\mu\nu}=(\zeta_{\mu}\lambda_{\nu}+\zeta_{\nu}\lambda_{\mu})/2$ where $\zeta$ and $\lambda$ are two vectors, and similarly the totally antisymmetry of the R-R three-form potential can be implemented by writing it as Slater determinant of three vectors.   Then one may transform  the couplings to the momentum space to find the independent ones. Using the fact that we are not interested in the couplings which can be absorbed by field redefinition of the supergravity, \ie  the couplings involve the Ricci and scalar curvatures and the divergence of the R-R field strength,  one can use the on-shell relations $k_i\inn k_i=0$ and $k_i\inn \epsilon_i=0$ where $\epsilon_i$ stands for the polarization of the $i$-th field in the momentum space.  Using conservation of momentum and these on-shell relations, one observes that even the above two independent terms are identical. We set the coefficient of one of them to zero and fixed the other one by requiring that the S-matrix element should not  produce the terms which have zero or one Mandelstam variable $k_i\inn k_j$ \cite{Garousi:2012yr,Barreiro:2012aw}.

Using the ten-dimensional action \reef{F4R}, one can easily find the corresponding couplings for R-R five-form field strength. To this end, one uses  the dimensional reduction and finds  the couplings   which have no $y$-index. They are given by the nine-dimensional form of the Lagrangian \reef{F4R} . Under the T-duality \reef{TF} they produce  the couplings with structure $\eta^{yy}F^{(5)}_yF^{(5)}_yRR$. Extending the $y$-index to complete spacetime index and taking the symmetry factors into account, one finds the following couplings in which the R-R field strengths contract with each other at least once: 
\beqa
S &\supset&\frac{\gamma}{\kappa^2} \int d^{10}x\sqrt{-G}\bigg[\frac{1}{160} F_{p r s t u,q}^2 R_{h k m n}-\frac{1}{96} F_{p r s t u,q} F_{q r s t u,p} R_{h k m n}+\frac{1}{12} F_{m r s t u,n} F_{p r s t u,q} R_{h k p q}\nonumber\\&&-\frac{1}{6} F_{h p s t u,q} F_{p r s t u,q} R_{m n k r}+\frac{1}{3} F_{h p s t u,q} F_{q r s t u,p} R_{m n k r}+\frac{1}{2} \left(\frac{1}{3} F_{k p r t u,q}-F_{k q r t u,p}\right) F_{h p s t u,q} R_{m n r s}\nonumber\\&&+\left(\frac{1}{3} F_{h p s t u,q}-\frac{1}{3} F_{h q s t u,p}\right) F_{n r s t u,q} R_{m p k r}+\left(\frac{1}{3} F_{h n s t u,q}-\frac{1}{3} F_{h q s t u,n}\right) F_{p r s t u,q} R_{m p k r}\nonumber\\&&-\left(\frac{1}{3} F_{h p s t u,q}-\frac{1}{6} F_{h q s t u,p}\right) F_{q r s t u,n} R_{m p k r}-\left(\frac{1}{3} F_{h n s t u,q}-\frac{1}{6} F_{h q s t u,n}\right) F_{q r s t u,p} R_{m p k r}\nonumber\\&&-2 \left(-F_{k n r t u,p}+\frac{1}{3} F_{k p r t u,n}\right) F_{h p s t u,q} R_{m q r s}-4 \left(\frac{1}{2} F_{k n r t u,p}-\frac{1}{2} F_{k p r t u,n}\right) F_{h q s t u,p} R_{m q r s}\nonumber\\&&-\frac{2}{3} F_{h q s t u,p} F_{n r s t u,m} R_{p q k r}-\frac{2}{3} F_{k n r t u,m} F_{q r s t u,p} R_{p q h s}+\frac{1}{2} F_{k n r t u,p} F_{h m s t u,q} R_{p q r s}\nonumber\\&&+\left(-F_{k n r t u,p}+\frac{1}{2} F_{k p r t u,n}\right) F_{h q s t u,m} R_{p q r s}+2 F_{k n r t u,m} F_{h q s t u,p} R_{p q r s}\nonumber\\&&+\frac{1}{4} F_{h k q t u,p} F_{n r s t u,m} R_{p q r s}-\frac{3}{4} F_{h k p q u,s} F_{m n r t u,s} R_{p q r t}\bigg]R_{h k m n}\labell{F5R}
\eeqa
In this case it is impossible to construct the coupling in which the R-R five-form field strengths have no contraction with each other.  So the above couplings must be the correct couplings of two Riemann curvatures and two R-R five-form field strengths in the string frame. 

These couplings are in type IIB theory so they can be extended to S-duality invariant forms. One has to again move to the Einstein frame. The dilaton factor in the Einstein frame becomes $e^{-3\phi_0/2}$ which must be extended to the Eisenstein series $E_{3/2}$, the  Einstein frame Riemann curvatures  and the R-R five-form field strength are also  invariant under the $SL(2,R)$ transformation. The S-duality invariant couplings produces no new tree-level couplings other than the couplings \reef{F5R}. So T-duality on the S-duality invariant form of \reef{F5R} would not give any new couplings.

\section{$F^{(n)}F^{(n)}HH$ }

In this section we are going to find the couplings with structure $F^{(n)}F^{(n)}HH$  and some of the four R-R couplings which are related to them. We begin by finding the couplings with structure $F^{(3)}F^{(3)}HH$ in type IIB theory and then find all other couplings by using consistency with linear T-duality, as we have done in the previous section. 

One may try to find the  couplings  $F^{(3)}F^{(3)}HH$ by extending the $H^4$ couplings in \reef{S} to the S-duality invariant form, using \reef{SHH}. However, the result depend on which pair of $H^2$ extend to $\cH^T\cM_0\cH$. There are many different choices which result many  different forms for the $F^{(3)}F^{(3)}HH$ couplings. They all give, however,  identical result for $F^{(3)}F^{(3)}F^{(3)}F^{(3)}$ couplings. Moreover, there are different S-duality invariant couplings which do not contain  $H^4$ terms at all. The couplings with structure $\cH^T\cN\cH\cH^T\cN\cH$ 
are invariant under the S-duality. They contain only couplings with structure $F^{(3)}F^{(3)}HH$.  We are going to find the $F^{(3)}F^{(3)}HH$ couplings by using linear T-duality on the couplings we have found in the previous section. The result should be the same as the couplings one finds by making the action \reef{S} to be S-duality invariant.

 Using the dimensional reduction on the couplings \reef{F4R}, we find the couplings with structure $\eta^{yy}\eta^{yy}F^{(4)}_yF^{(4)}_yR_yR_y$. Under the T-duality they transforms to the couplings with structure $F^{(3)}F^{(3)}H_yH_y$. One may easily  find the couplings $F^{(3)}F^{(3)}HH$ in which the B-field strengths contract with each other. However, there are too many other terms in which the B-field strengths contract with the R-R three-form field strengths. On the other hand, the couplings $F^{(3)}F^{(3)}HH$ must satisfy in fact four  constraints:

1-They should produce the above couplings $F^{(3)}F^{(3)}H_yH_y$.

2-They should produce the couplings $F^{(3)}_yF^{(3)}_yHH$ which are the transformation of the couplings $F^{(3)}F^{(3)}H_yH_y$ under the S-duality.

3-They should produce the couplings $F^{(3)}_yF^{(3)}_yH_yH_y$ which are the transformation of the couplings in \reef{F2R} with structure $F^{(2)}F^{(2)}R_yR_y$ under the T-duality.

4-They should produce the couplings $F^{(3)}_yF^{(3)}HH_y$ which are the transformation of the couplings  with structure $F^{(2)}F^{(4)}_yHR_y$ under the T-duality. We have found the couplings with structure  $F^{(2)}F^{(4)}HR$ in the Appendix from T-duality of the couplings \reef{F3R}. 

%One way to find the  couplings $F^{(3)}F^{(3)}HH$ is to convert the $y$-index in each of the above terms to a complet spacetime index and to consider the coefficient of that term with an unknown constant. 
One may consider the couplings  $F^{(3)}F^{(3)}HH$ with all possible contractions of the indices and choose an unknown coefficient for each of them. Then one should use the dimensional reduction on the resulting couplings and find the couplings with structure $F^{(3)}F^{(3)}H_yH_y$, $F^{(3)}_yF^{(3)}_yHH$, $F^{(3)}_yF^{(3)}_yH_yH_y$ and $F^{(3)}_yF^{(3)}HH_y$. Then one can find  their coefficients by forcing them to satisfy the above four constraints.  One can use the Bianchi identity by writing the couplings  in terms of potential \ie $F=\frac{1}{2}dC$ and $H=\frac{1}{2}dB$, and transforming the couplings to the momentum space. We have found that the constraint equations are not consistent with each other. In particular, the couplings can not satisfy the  constraints 1 and 3 at the same time. We will discuss the reason for this inconsistency in the discussion section.  However, if one uses the on-shell relations $k_i\inn k_i=0$ and $k_i\inn \eps_i=0$, the number of independent variables decreases, \eg $k_1\inn k_2=k_3\inn k_4$. This makes, for example,   the  couplings in the constraint 4 to be  zero.   Using the on-shell relations, one finds that the above constraint equations  become  consistent and give consistent relations between the unknown coefficients.  Solving these equations one finds a set of couplings which  have still some unknown coefficients. We then impose the condition that the S-matrix does not produce couplings with zero and one Mandelstam variable $k_i\inn k_j$. This gives one extra relation between the coefficients. Using this relation, one finds interestingly a set of couplings in terms of the potentials $B,C$  which have no  unknown coefficients any more! That means  if we consider the original couplings which are in terms of field strengths $H,F$, and use the relations between the coefficients that we have found, then the left over constants has to cancel themselves  using the Bianchi identity and the on-shell relations.  Therefore, the left over constant can be set all to zero. We have found the following result in the string frame:
\beqa
S &\supset&\frac{\gamma}{\kappa^2} \int d^{10}x\sqrt{-G}\bigg[-8 F_{h r s,n} F_{q r s,m} H_{k n p,h} H_{m p q,k}+8 F_{n q s,h} F_{p q s,m} H_{k n r,h} H_{m p r,k}\nonumber\\&&-\frac{2}{9} F_{n q s,h} F_{n q s,k} H_{m p r,h} H_{m p r,k}+2 F_{h q s,n} F_{k q s,m} H_{m p r,k} H_{n p r,h}+2 F_{h q s,m} F_{k q s,n} H_{m p r,k} H_{n p r,h}\nonumber\\&&+2 F_{n q r,h} F_{n p s,k} H_{h q r,m} H_{k p s,m}-2 F_{n q r,h} F_{m r s,k} H_{n p q,k} H_{m p s,h}-8 F_{m n p,k} F_{n r s,h} H_{k p q,m} H_{h q s,r}\nonumber\\&&+8 F_{n q r,h} F_{m p s,k} H_{h k r,n} H_{p q s,m}+2 F_{m n p,h} F_{m q s,h} H_{n p r,k} H_{q r s,k}\bigg]\labell{F3H}
\eeqa 
Since we have used the on-shell relations, there are many other couplings which involve the divergence of $H$ and $F$ in which we are not interested, because  they can be absorbed in the supergravity by appropriate field redefinition.

To study the S-duality of the above couplings, one has to go to the Einstein frame. In this frame,  one finds the overall dilaton factor to be $e^{-3\phi_0/2}$ which is extended to $E_{3/2}$ in S-duality invariant form.  Writing the above couplings in terms of potentials $B,C$, one observes that they are invariant under  $B\rightarrow C,\, C\rightarrow -B$ which is consistent with S-duality. However, as we mentioned before at the beginning of this section, there are ambiguities in writing the couplings in  $SL(2,R)$ invariant form. We leave this part for the future works.

Having found the couplings with structure $F^{(3)}F^{(3)}HH$, one can now find the couplings  with structure $F^{(2)}F^{(2)}HH$ in type IIA theory.  Under the dimensional reduction on the couplings \reef{F3H}, one should first find  the couplings with structure  $\eta^{yy}F^{(3)}_yF^{(3)}_yHH$   and then use the T-duality rule \reef{TF}. The result is
\beqa
S &\supset&\frac{\gamma}{\kappa^2} \int d^{10}x\sqrt{-G}\bigg[-16 F_{h r,n} F_{q r,m} H_{k n p,h} H_{m p q,k}+16 F_{n q,h} F_{p q,m} H_{k n r,h} H_{m p r,k}\nonumber\\&&-\frac{2}{3} F_{n q,h} F_{n q,k} H_{m p r,h} H_{m p r,k}+4 F_{h q,n} F_{k q,m} H_{m p r,k} H_{n p r,h}+4 F_{h q,m} F_{k q,n} H_{m p r,k} H_{n p r,h}\nonumber\\&&+2 F_{q r,h} F_{p s,k} H_{h q r,m} H_{k p s,m}+2 F_{n q,h} F_{m s,k} H_{n p q,k} H_{m p s,h}+8 F_{m p,k} F_{r s,h} H_{k p q,m} H_{h q s,r}\nonumber\\&&+2 F_{n p,h} F_{q s,h} H_{n p r,k} H_{q r s,k}\bigg]\labell{F2H}
\eeqa
Here again the indices run over all directions except the $y$-direction and the fields are independent of the $y$-direction. After using the T-duality, one  assumes the fields depend on all directions and the indices run over all spacetime directions. We have checked that the Fourier transform of the above couplings to the momentum space has no couplings with zero and one $k_i\inn k_j$ after using on-shell relations.

The couplings  with structure $F^{(1)}F^{(1)}HH$   can  be found from the above couplings by using the dimensional reduction and finding the terms with structure $\eta^{yy}F^{(2)}_yF^{(2)}_yHH$. Then under  T-duality they produce the following terms in the string frame:
\beqa
S &\supset&\frac{\gamma}{\kappa^2} \int d^{10}x\sqrt{-G}\bigg[-\frac{4}{3} F_{n,h} F_{n,k} H_{m p r,h} H_{m p r,k}+4 F_{h,n} F_{k,m} H_{m p r,k} H_{n p r,h}\nonumber\\&&+4 F_{h,m} F_{k,n} H_{m p r,k} H_{n p r,h}\bigg]\labell{F1H}
\eeqa
In the Einstein frame they  can be extended to the following S-duality invariant form:
\beqa
S &\supset&\frac{\gamma}{\kappa^2} \int d^{10}x\sqrt{-G}E_{3/2}\bigg[\frac{1}{6} \Tr[\cM_{,nh}\cM^{-1}_{,nk} ] \cH^T_{m p r,h}\cM_0 \cH_{m p r,k}\labell{F1HS}\\&&-\frac{1}{2} \Tr[\cM_{,nh}\cM^{-1}_{,km} ] \cH^T_{m p r,k} \cM_0\cH_{n p r,h}-\frac{1}{2} \Tr[\cM_{,hm}\cM^{-1}_{,kn} ] \cH^T_{m p r,k} \cM_0\cH_{n p r,h}\bigg]\nonumber
\eeqa
The above action contains the following tree-level couplings of two B-field strengths  and two dilatons in the string frame:
\beqa
S &\supset&\frac{\gamma}{\kappa^2} \int d^{10}x\sqrt{-G}e^{-2\phi_0}\kappa^2\bigg[-\frac{2}{3} \phi_{,nh} \phi_{,nk} H_{m p r,h} H_{m p r,k}+2 \phi_{,hn} \phi_{,km} H_{m p r,k} H_{n p r,h}\nonumber\\&&+2 \phi_{,hm} \phi_{,kn} H_{m p r,k} H_{n p r,h}\bigg]
\eeqa
Using the on-shell relations, we have checked that the above  couplings are exactly the $H^2\phi^2$ terms  in \reef{S}. So the $H^2\phi^2$ terms  in \reef{S} and  the couplings \reef{F1H} are two components of one  $SL(2,R)$ invariant set. This set  \reef{F1HS} at zero axion background, \ie $C_0=0$,  includes also the couplings with structures $F^{(1)}F^{(1)}F^{(3)}F^{(3)}$ and $\phi\phi F^{(3)}F^{(3)}$ at the tree-level. To apply T-duality on \reef{F1HS} one has to move back to the string frame in which the latter couplings and the $H^2\phi^2$ terms  in \reef{S} do not survive. It is difficult to find new couplings by applying T-duality on the couplings $F^{(1)}F^{(1)}F^{(3)}F^{(3)}$ in the string frame to find the couplings between  higher R-R forms. We will find them later on by some other means.

To find the couplings  $F^{(4)}F^{(4)}HH$ in type IIA theory, we first consider $F^{(3)}F^{(3)}HH$ couplings in \reef{F3H} which carry no Killing index. Under T-duality they produce the couplings with structure  $F^{(4)}_yF^{(4)}_yHH$. Using  these couplings, one can easily find the couplings in which the R-R field strengths contract with each other. They are given by the couplings in the first five lines of the following equation:
\beqa
S &\supset&\frac{\gamma}{\kappa^2} \int d^{10}x\sqrt{-G}\bigg[-\frac{8}{3} F_{h r s t,n} F_{q r s t,m} H_{k n p,h} H_{m p q,k}+\frac{8}{3} F_{n q s t,h} F_{p q s t,m} H_{k n r,h} H_{m p r,k}\nonumber\\&&-\frac{1}{18} F_{n q s t,h} F_{n q s t,k} H_{m p r,h} H_{m p r,k}+\frac{2}{3} F_{h q s t,n} F_{k q s t,m} H_{m p r,k} H_{n p r,h}\nonumber\\&&+\frac{2}{3} F_{h q s t,m} F_{k q s t,n} H_{m p r,k} H_{n p r,h}+F_{m n p t,h} F_{m q s t,h} H_{n p r,k} H_{q r s,k}\nonumber\\&&+F_{n q r t,h} F_{n p s t,k} H_{h q r,m} H_{k p s,m}-F_{n q r t,h} F_{m r s t,k} H_{n p q,k} H_{m p s,h}\nonumber\\&&-4 F_{m n p t,k} F_{n r s t,h} H_{k p q,m} H_{h q s,r}+8 F_{n q r t,h} F_{m p s t,k} H_{h k r,n} H_{p q s,m}\nonumber\\&&+\frac{4}{9} F_{h k m n,t} F_{p q r t,n} H_{h k m,s} H_{p q r,s}\bigg]\labell{F4H}
\eeqa
The couplings in the last line has the structure that the R-R field strengths do not contract with each other. Using the Bianchi identity, one finds there are two such couplings, \ie
$b1 F_{h k m n,t} F_{p q r s,t} H_{h k p,q} H_{m n r,s}+b2 F_{h k m n,t} F_{p q r t,n} H_{h k m,s} H_{p q r,s}$. The unknown constants $b1,b2$ can be found by using the fact that S-matrix does not produce couplings with zero and one $k_i\inn k_j$. This fixes these constants to be $b1=0$ and $b2=4/9$.

Under the dimensional reduction, the couplings $F^{(4)}F^{(4)}HH$ which have no $y$-index are given by \reef{F4H}. Under the T-duality they produce correctly the couplings  $\eta^{yy}F^{(5)}_yF^{(5)}_yHH$. Completing the $y$-index and taking into account the symmetry factors, one can easily find the couplings in which the R-R field strength contract with each other at least once. In the string frame, they are
\beqa
S &\supset&\frac{\gamma}{\kappa^2} \int d^{10}x\sqrt{-G}\bigg[-\frac{2}{3} F_{h r s t u,n} F_{q r s t u,m} H_{k n p,h} H_{m p q,k}+\frac{2}{3} F_{n q s t u,h} F_{p q s t u,m} H_{k n r,h} H_{m p r,k}\nonumber\\&&-\frac{1}{90} F_{n q s t u,h} F_{n q s t u,k} H_{m p r,h} H_{m p r,k}+\frac{1}{6} F_{h q s t u,n} F_{k q s t u,m} H_{m p r,k} H_{n p r,h}\nonumber\\&&+\frac{1}{6} F_{h q s t u,m} F_{k q s t u,n} H_{m p r,k} H_{n p r,h}+\frac{4}{9} F_{h k m n u,t} F_{p q r t u,n} H_{h k m,s} H_{p q r,s}\nonumber\\&&+\frac{1}{3} F_{n q r t u,h} F_{n p s t u,k} H_{h q r,m} H_{k p s,m}-\frac{1}{3} F_{n q r t u,h} F_{m r s t u,k} H_{n p q,k} H_{m p s,h}\nonumber\\&&-\frac{4}{3} F_{m n p t u,k} F_{n r s t u,h} H_{k p q,m} H_{h q s,r}+4 F_{n q r t u,h} F_{m p s t u,k} H_{h k r,n} H_{p q s,m}\nonumber\\&&+\frac{1}{3} F_{m n p t u,h} F_{m q s t u,h} H_{n p r,k} H_{q r s,k}\bigg]\labell{F5H}
\eeqa
There is no coupling in which  the two R-R five-forms have no contraction with each other.  

To extend the above couplings to the S-duality invariant form, we first transform the couplings to the Einstein frame in which the dilaton factor  becomes $e^{-5\phi_0/2}$. This can be written as $e^{-3\phi_0/2}\times e^{-\phi_0/2}$. The first factor must be extended to the Eisenstein series $E_{3/2}$ and the second factor combined with the B-field strengths must be  written  as $SL(2,R)$ invariant form of $\cH^T\cM_0\cH$. The R-R five-form is also invariant under the S-duality. The S-duality invariant couplings at $C_0=0$ produce the couplings \reef{F5H}, as well as the similar couplings with structure $F^{(5)}F^{(5)}F^{(3)}F^{(3)}$.

 One can use another T-duality on the S-duality invariant  form of \reef{F5H} to find  various new couplings between four R-R field strengths by performing the following simple steps. One can use the dimensional reduction on the  above $F^{(5)}F^{(5)}F^{(3)}F^{(3)}$ couplings to find the couplings with structure $F^{(5)}_yF^{(5)}_yF^{(3)}_yF^{(3)}_y$. Under T-duality \reef{TF} they then transform to  the couplings with structure $F^{(4)}F^{(4)}F^{(2)}F^{(2)}$ without any ambiguity. Using the resulting couplings one may use the dimensional reduction to find the couplings with structure $F^{(4)}F^{(4)}F^{(2)}_yF^{(2)}_y$. Then under T-duality they transform to the couplings $F^{(5)}_yF^{(5)}_yF^{(1)}F^{(1)}$. Converting the $y$-index to a complete spacetime index and taking the symmetry factors into account, one finds the couplings with structure $F^{(5)}F^{(5)}F^{(1)}F^{(1)}$ without any ambiguity because it is impossible to have couplings in which the R-R five-forms do not contract with each other. Using the S-duality invariant extension of the two R-R one-form field strengths \reef{FF}, one can then write the resulting couplings in an  S-duality invariant form by adding the appropriate $F^{(5)}F^{(5)}\phi\phi$ couplings.

Under the dimensional reduction of the couplings $F^{(5)}F^{(5)}F^{(3)}F^{(3)}$, one may also consider the couplings with structure $F^{(5)}_yF^{(5)}_yF^{(3)}F^{(3)}$. Under T-duality \reef{TF} they transform to the couplings with structure $F^{(4)}F^{(4)}F^{(4)}_yF^{(4)}_y$. Then by converting the $y$-index to complete spacetime index and taking   the symmetry factors into account, one can find the couplings with structure  $F^{(4)}F^{(4)}F^{(4)}F^{(4)}$ without ambiguity because  all the R-R field strengths are identical, as a result,  there would be no term in which the two R-R  field strengths do not contract with each other.  

The couplings with structure $F^{(3)}F^{(3)}F^{(3)}F^{(3)}$ can be found either by using T-duality on $F^{(4)}_yF^{(4)}_yF^{(4)}_yF^{(4)}_y$, or by making the $H^4$ couplings in \reef{S}  to be S-duality invariant. We already mentioned that it is ambitious in finding the $SL(2,R)$ invariant form of the $H^4$ couplings, however, there is no ambiguity in finding $F^{(3)}F^{(3)}F^{(3)}F^{(3)}$ terms from $H^4$. Using these  $F^{(3)}F^{(3)}F^{(3)}F^{(3)}$ couplings, one can easily find the couplings with structure  $F^{(2)}F^{(2)}F^{(2)}F^{(2)}$ and $F^{(1)}F^{(1)}F^{(1)}F^{(1)}$. The latter couplings  should be combined with  the $\phi^4$ couplings in action \reef{S} and some $F^{(1)}F^{(1)}\phi\phi$ couplings to become invariant under the $SL(2,R)$ transformations.

\section{$F^{(n)}F^{(n+2)}HR$}

Using the dimensional reduction on the couplings \reef{F4H} in type IIA theory, one can find the couplings with structure $\sqrt{-G}\eta^{yy}F^{(4)}_yF^{(4)}HH_y$. Under the T-duality transformations \reef{TR} and \reef{TF}, they transform to the couplings with structure $\sqrt{-G}\eta^{yy}\eta_{yy}F^{(3)}F^{(5)}_yH\eta^{yy}R_y$ in type IIB theory. They produce the couplings in which the five-form field strength contract with the Riemann curvature. They appear in the following action in the string frame:
\beqa
S &\!\!\!\!\!\supset\!\!\!\!\!&\frac{\gamma}{\kappa^2} \int d^{10}x\sqrt{-G}\bigg[ 8 F_{n q t,h} F_{m p r s t,k} H_{p q s,m} R_{h k n r}+\frac{8}{3} F_{r s t,m} F_{h q r s t,n} H_{k n p,h} R_{m p k q}\labell{F3F5}\\&&-8 F_{m n t,k} F_{n p r s t,h} H_{h q s,r} R_{k q m p}-\frac{8}{3} F_{q s t,h} F_{n p q s t,m} H_{m p r,k} R_{k r h n}+\frac{8}{3} F_{q s t,m} F_{n p q s t,h} H_{k n r,h} R_{m r k p}\nonumber\\&&-\frac{8}{3} F_{h k m n p,t} F_{q r t,n} H_{h k m,s} R_{q r p s}-4 F_{n p q r t,h} F_{n s t,k} H_{h q r,m} R_{k s m p}-4 F_{m n q r t,h} F_{r s t,k} H_{n p q,k} R_{p s h m}\nonumber\\&&+8 F_{n p q r t,h} F_{m s t,k} H_{h k r,n} R_{q s m p}+4 F_{m n p q t,h} F_{m s t,h} H_{n p r,k} R_{r s k q}-4 F_{h k m n p,t} F_{h k r,m} H_{n p q,s} R_{q r s t}\bigg]\nonumber
\eeqa
The last term however, is a coupling which has no contraction between the five-form field strength and  the Riemann curvature. There are some other such couplings. Without these terms the couplings are not consistent with S-duality. Since we are not interested in the Ricci curvature and the divergence of the B-field and the R-R field strengths, one may find the coefficients of such terms by imposing both the Bianchi identity and the on-shell relations. We have found that  by adding the last term and using the on-shell relations, the couplings can be written in the $SL(2,R)$ invariant form of $\cH^T\cN\cH$. 

In the Einstein frame, the dilaton factor  is $e^{-3\phi_0/2}$ which is extended to $E_{3/2}$, the  Riemann curvature and the R-R field strength are also invariant under the S-duality. The S-duality invariant couplings then produce only the couplings \reef{F3F5} at  tree level. So the T-duality on them does not produce any new couplings. We have  checked that the above couplings  satisfy the condition that the couplings in the momentum space have no zero or one Mandelstam variable $k_i\inn k_j$.

The couplings with structure $F^{(2)}F^{(4)}HR$ in type IIA theory  can be found from dimensional reduction of the above couplings. To this end, one should first find the couplings with structure $\sqrt{-G}\eta^{yy}F^{(3)}_yF^{(5)}_yHR$. Then under T-duality they transform to the couplings with structure  $\sqrt{-G}F^{(2)}F^{(4)}HR$. The result in the string frame is
\beqa
S &\supset&\frac{\gamma}{\kappa^2} \int d^{10}x\sqrt{-G}\bigg[8 F_{n q,h} F_{m p r s,k} H_{p q s,m} R_{h k n r}+8 F_{r s,m} F_{h q r s,n} H_{k n p,h} R_{m p k q}\nonumber\\&&-16 F_{m n,k} F_{n p r s,h} H_{h q s,r} R_{k q m p}-8 F_{q s,h} F_{n p q s,m} H_{m p r,k} R_{k r h n}+8 F_{q s,m} F_{n p q s,h} H_{k n r,h} R_{m r k p}\nonumber\\&&+8 F_{h m n p,k} F_{h r,m} H_{n p q,s} R_{q r k s}-8 F_{n p q r,h} F_{n s,k} H_{h q r,m} R_{k s m p}-8 F_{m n q r,h} F_{r s,k} H_{n p q,k} R_{p s h m}\nonumber\\&&+8 F_{n p q r,h} F_{m s,k} H_{h k r,n} R_{q s m p}+8 F_{m n p q,h} F_{m s,h} H_{n p r,k} R_{r s k q}\bigg]\labell{F2F4}
\eeqa
There are no extra terms that has to be added. We have checked that the S-matrix element of the above couplings produces no term with zero and one factor of $k_i\inn k_j$, as expected. There is a different representation for the couplings with structure $F^{(2)}F^{(4)}HR$ in the Appendix which has been found from T-duality of the couplings \reef{F3R}. That representation and the above representation are identical using the on-shell relations. So the difference between them are some terms which can be absorbed into supergravity by field redefinition.

Doing the same steps as above, one  finds the  couplings with structure $F^{(1)}F^{(3)}HR$ in type IIB theory. In the string frame, they are  the following:
\beqa
S &\supset&\frac{\gamma}{\kappa^2} \int d^{10}x\sqrt{-G}\bigg[ 16 F_{r,m} F_{h q r,n} H_{k n p,h} R_{m p k q}-16 F_{m,k} F_{n p r,h} H_{h n q,r} R_{k q m p}\nonumber\\&&-16 F_{q,h} F_{n p q,m} H_{m p r,k} R_{k r h n}+16 F_{q,m} F_{n p q,h} H_{k n r,h} R_{m r k p}-8 F_{m,h} F_{n p q,h} H_{n p r,k} R_{m r k q}\nonumber\\&&+8 F_{m n q,h} F_{r,k} H_{n p q,k} R_{p r h m}-8 F_{m n p,k} F_{r,m} H_{n p q,h} R_{q r h k}\bigg]\labell{F1F3}
\eeqa
In the Einstein frame, the dilaton factor is $e^{-\phi_0/2}$. This can be rewritten as $e^{-3\phi_0/2}\times e^{\phi_0}$. The first factor again is extended to $E_{3/2}$. The second factor combined with the R-R one-form field strength, transforms as
\beqa
F^{(1)}e^{\phi_0}\rightarrow -F^{(1)}e^{\phi_0}
\eeqa
in zero axion background, $C_0=0$,  and at the linear order of quantum states. On the other hand, using on-shell relations we have checked that the amplitude is antisymmetric under the transformation 
\beqa
F^{(3)}\rightarrow H&;& H\rightarrow -F^{(3)}
\eeqa
Therefore, the action \reef{F1F3} is consistent with  the S-duality for the zero axion background. The S-duality invariant extension of the above couplings for nonzero $C_0$  are
\beqa
S &\supset&\frac{\gamma}{\kappa^2} \int d^{10}x\sqrt{-G}\, E_{3/2}\bigg[ 4\cH^T_{h q r,n}\cM_{,rm}\cH_{k n p,h} R_{m p k q} +\cdots\bigg]\labell{F1F31}
\eeqa
where dotes refer to the other terms in \reef{F1F3}. 
For zero axion background, one finds
\beqa
\cH^T_{h q r,n}\cM_{,rm}\cH_{k n p,h}R_{m p k q}&=&2e^{\phi_0}F_{r,m}(F_{h q r,n}H_{k n p,h}+H_{h q r,n}F_{k n p,h})R_{m p k q}\nonumber\\
&&+\sqrt{2}\kappa\phi_{,rm}(e^{\phi_0}F_{h q r,n}F_{k n p,h}-e^{-\phi_0}H_{h q r,n}H_{k n p,h})R_{m p k q}
\eeqa
 Therefore,  the S-duality invariant action \reef{F1F31} produces the couplings  \reef{F1F3} after  using the   on-shell relations. It also    produces the couplings between one Riemann curvature, two two-form field strengths and one dilaton. The $HHR\phi$ couplings in \reef{F1F31} should be  the same as the corresponding couplings in   \reef{S} after using the on-shell relations. In other words, the  $HHR\phi$ couplings in \reef{S} and the couplings \reef{F1F3} are two components of the above $SL(2,R)$ invariant.

\section{$F^{(n)}F^{(n+4)}RR$}

The R-R fields in the supergravity are $C^{(0)},\cdots , C^{(4)}$. So there is only one possible value for $n$ which is $n=1$.   To find $F^{(1)}F^{(5)}RR$ couplings, one has to use dimensional reduction on the couplings \reef{F2F4} and find the couplings with structure  $\eta^{yy}F^{(2)}_yF^{(4)}H_yR$. There is only one term which is
\beqa
-8 \eta^{yy}R_{h k n r} F_{m p r s,k} F_{n y,h} H_{p s y,m}\nonumber
\eeqa
Under T-duality it becomes 
\beqa
8 R_{h k n r} F_{m p r s y,k} F_{n,h} R_{p s m y}\nonumber
\eeqa
Completing the $y$-index, one finds the coupling $R_{h k n r} F_{m p r s q,k} F_{n,h} R_{p s m q}$ which is zero using the antisymmetry of the R-R five-form and the symmetry of the Riemann curvature. So our calculation indicates that there is no coupling with structure $F^{(1)}F^{(5)}RR$. This is what one expects because the supergravity does not produce massless pole for the scattering amplitude of one R-R scalar, one R-R four-form and two gravitons.  So the corresponding S-matrix element in string theory is zero. As a result, there is no coupling $F^{(1)}F^{(5)}RR$ at order $\alpha'^3$.

\section{$F^{(n)}F^{(n+4)}HH$}

To find $F^{(1)}F^{(5)}HH$ couplings, one has to use dimensional reduction on the couplings \reef{F2F4} and find the couplings with structure  $\eta^{yy}F^{(2)}_yF^{(4)}HR_y$. Then under T-duality they transform to the couplings with structure $F^{(1)}F^{(5)}_yHH_y$. Using them one can easily find the couplings  $F^{(1)}F^{(5)}HH$ in which the R-R five-form contracts with $H$. In the string frame, they are given by
\beqa
S &\supset&\frac{\gamma}{\kappa^2} \int d^{10}x\sqrt{-G}\bigg[ -8 F_{n,k} F_{n p q r s,h} H_{h q r,m} H_{m p s,k}+16 F_{n,k} F_{m n p r s,h} H_{k m q,p} H_{h q s,r}\nonumber\\&&+4 F_{m,h} F_{m n p q s,h} H_{n p r,k} H_{k q s,r}-16 F_{q,h} F_{m n p r s,k} H_{h k n,r} H_{p q s,m}\bigg]\labell{F1F5}
\eeqa
There is no possibility for the R-R five-form to have no contraction with $H$, so the above couplings must be the desired couplings. 

In the Einstein frame, the overall dilaton factor is $e^{-3\phi_0/2}$ which is extended to $E_{3/2}$ in the S-duality invariant couplings. The terms in the bracket must be extended to $SL(2,R)$ invariant forms. The S-duality invariant form of the above couplings are
\beqa
S &\supset&\frac{\gamma}{\kappa^2} \int d^{10}x\sqrt{-G}\, E_{3/2}\bigg[ -4 \cH^T_{h q r,m}\cM_{,nk}\cN^{-1}\cM_0\cH_{m p s,k} F_{n p q r s,h} +\cdots\bigg]\labell{F1F51}
\eeqa
where dots refer to similar constructions for other terms of \reef{F1F5}. 
For zero axion background, one finds
\beqa
\cH^T_{h q r,m}\cM_{,nk}\cN^{-1}\cM_0\cH_{m p s,k}&=&2F_{n,k}(H_{h q r,m} H_{m p s,k}-e^{2\phi_0}F_{h q r,m} F_{m p s,k})\nonumber\\
&&+\sqrt{2}\kappa\phi_{,nk}(H_{h q r,m} F_{m p s,k}+F_{h q r,m} H_{m p s,k})
\eeqa
So the S-duality invariant action \reef{F1F51} produces   the couplings in \reef{F1F5} and the couplings with structure $F^{(1)}F^{(5)}F^{(3)}F^{(3)}$. The terms corresponding to the second line above have structure $F^{(5)}F^{(3)}H\phi$. In the string frame these couplings should not survive. In fact, when  transforming the Riemann curvature from Einstein frame to the string frame, one finds the second derivative of the dilaton. In transforming the Einstein frame couplings \reef{F3F5} to the string frame, one would find the above $F^{(5)}F^{(3)}H\phi$ couplings with minus sign.  

The non-zero couplings \reef{F1F51} are also consistent with the fact that the supergravity at order $\alpha'^0$ produces massless poles for the scattering of  one R-R scalar, one R-R four-form and two B-fields.  So the corresponding S-matrix element in string theory is not zero. It should produce the couplings \reef{F1F51}  at order $\alpha'^3$.

\section{Discussion}

In this paper we have studied the invariance  of the NS-NS corrections to the supergravity found by Gross and Sloan \cite{Gross:1986mw} under S-duality and T-duality  to find new couplings in the supergravity at order $\alpha'^3$ which involve R-R fields.  The new couplings satisfy these dualities after using on-shell relations in the momentum space, \ie $k_i\inn k_i=0$ and $k_i\inn\eps_i=0$ where $\eps_i$ is the polarization of the fields. The Ricci and scalar curvatures as well as the divergence of the form-field strengths produce terms which have $k_i\inn k_i$ and $k_i\inn\eps_i$ in the momentum space. So using the on-shell relations means the corrections that we have found are off by the couplings involving the Ricci and scalar curvatures as well as the divergence of the form-field strengths. On the other hand, these couplings can be absorbed into the supergravity at order $\alpha'^0$ by field redefinition \cite{Gross:1986iv}. Therefore, the couplings  should be valid up to field redefinitions.

We have seen that the couplings with structure $F^{(1)}F^{(1)}RR$ in  \reef{F1R} which are coming from the T-duality of the couplings with structure $F^{(2)}F^{(2)}RR$ in \reef{F2R}, and the couplings with structure $\phi\phi RR$  in \reef{F1R1} which are coming from the ten-dimensional action \reef{S}, combine together to form the S-duality invariant action \reef{F1R2}. This indicates that there is no other  $\phi\phi RR$ couplings than those in the ten-dimensional action \reef{S}.  This invalidates the interpretation  in \cite{Garousi:2012jp} which considers the Lagrangian $\cL$  to be  eight-dimensional and adds  some dilaton couplings to extend  it to the ten-dimensional Lagrangian.  The action \reef{S} in the string frame has no dilaton other than the overall factor $e^{-2\phi_0}$. We expect similar thing happens for  all  other couplings.

In the string frame, we have found   all couplings involving  the  R-R field strengths, except the couplings of four R-R five-form field strengths. These couplings which are invariant under the S-duality are related by the  T-duality to the couplings of four R-R four-form field strengths which have been found in section 5.  The T-duality of latter couplings gives couplings  with structure $F^{(5)}_yF^{(5)}_yF^{(5)}_yF^{(5)}_y$. There is no other constraint. In this case, however, there is no unique way to complete the $y$-indices.   One way to find the  $F^{(5)}F^{(5)}F^{(5)}F^{(5)}$ couplings is to consider  all possible contractions of the five-forms   and  to choose an unknown coefficient for each of them. Then one can   find some relations between the coefficients by using the dimensional reduction on the couplings  and comparing the result with the above $F^{(5)}_yF^{(5)}_yF^{(5)}_yF^{(5)}_y$ terms.  Using the Bianchi identity, the on-shell relations and the constraint that the S-matrix should have no term with zero or one Mandelstam variable  \cite{Garousi:2012yr,Barreiro:2012aw}, one may be able to fix all the coefficients, as we have done in section 5. It would be interesting to perform this calculation.

We have found the R-R couplings by requiring the compatibility of the NS-NS couplings \reef{S} with S-duality and T-duality. These couplings can be compared with the direct on-shell string theory calculations. One may  confirm these couplings by   direct   sphere-level S-matrix calculations in the RNS formalism.  In type IIB theory, the couplings at sphere-level and one-loop level are identical up to the overall dilaton factor. So one can compare the couplings in type IIB theory with the one-loop calculations which have be done in \cite{Peeters:2003pv} for  the couplings of two R-R field strengths and two Riemann curvatures.  The R-R couplings in both type IIA and IIB superstring theories, have been also found in \cite{Policastro:2006vt} using the pure-spinor formalism. The results are in terms of the trace of  the Gamma-matrices so one needs a symbolic program like \cite{Gran:2001yh}  for manipulating the Gamma-matrices to find the explicit form of the couplings. It would be interesting to compare the explicit couplings   in the present paper with the results in \cite{Peeters:2003pv,Policastro:2006vt}.

We have extended the action \reef{S} to be invariant under sequence of S-duality and T-duality, up to  on-shell relations. To implement the on-shell relations, one has to use the conservation of momentum in the momentum space or has to use the  integration by parts in the spacetime. Therefore, the Lagrangian has three parts. One part contains  the couplings that involve the Riemann curvatures and the derivative of the form-field strengths that we have found in this paper. Another part contains couplings which involve the Ricci and scalar curvatures as well as the divergence of the form-field strengths. This part which   may also be invariant under S-duality and T-duality, can be absorbed into the supergravity at order $\alpha'^0$. The last part contains total derivative terms. This part  which has no effect in the action, can not be invariant under the T-duality and the S-duality  because the T-duality transforms the curvature to derivative of $H$, the S-duality changes the derivative of $H$ to the derivative of the R-R three-form field strength, and another T-duality transforms this to the other R-R field strengths. So the invariant terms must have only derivative of field strengths. So the total derivative terms can not be invariant under the T- and S-dualities.  However, one may expect the Lagrangian itself to be invariant under these dualities. In this case, one should find a Lagrangian which is invariant under the  S-duality and T-duality  without using the conservation of momentum. 

It has been observed in \cite{Garousi:2012jp} that the Lagrangian $\cL$ in \reef{S} is invariant under the linear T-duality without using any the conservation of momentum. Our present calculation however indicates that the Lagrangian $\cL$ is not consistent with the S-duality and the T-duality without using conservation of momentum and the on-shell relations. That may indicate that the Lagrangian $\cL$ in \reef{S}  contains another set of Riemann curvatures which is zero on-shell. For example,  the following set of couplings has non-vanishing contribution at four gravitons level but is zero on-shell: 
\beqa
&& 12 R_{h k m n} R_{h p r s} R_{k q r s} R_{p q m n}-96 R_{h m k n} R_{m p n q} R_{h r p s} R_{k r q s}-24 R_{h m n p} R_{k m n p} R_{h r q s} R_{k r q s}\nonumber\\&&+24 R_{h m k n} R_{m p n q} R_{h r k s} R_{p r q s}+48 R_{h n k m} R_{m p n q} R_{h r p s} R_{k s q r}+3 R_{h k m n} R_{m n p q} R_{p q r s} R_{r s h k}\nonumber\\&&+\frac{3}{2} R_{h k m n}^2 R_{r s p q}^2
\eeqa
At the four gravitons level, the above couplings are identical to the Ricci and scalar curvatures. This is not true, however, at five and more gravitons levels. The consistency of the Lagrangian with  S-duality and T-duality may require adding the above couplings to the Lagrangian $\cL$ in \reef{S}. It would be interesting to perform this calculation to find a Lagrangian density which would be invariant under sequence of S-duality and T-duality without using integration by parts.

  {\bf Acknowledgments}:    This work is supported by Ferdowsi University of Mashhad under grant 2/26232-1392/02/11.

\newpage

{\bf\Large Appendix: $F^{(2)}F^{(4)}HR$ }

In this appendix we are going to find $F^{(2)}F^{(4)}RH$ couplings from the couplings \reef{F3R}. To this end, we first use dimensional reduction on the couplings \reef{F3R} to find the couplings with structure $F^{(3)}_yF^{(3)}RR_y$. Under the linear T-duality, they transforms to the couplings with structure $F^{(2)}F^{(4)}_yRH_y$. Inspired by these couplings one can find the couplings $F^{(2)}F^{(4)}RH$ in which the R-R four-forms are  contracted with $H$. Using the Bianchi identity, one finds that there are two couplings in which the R-R four-form is not contracted with $H$. Adding these two terms with unknown coefficient, one finds the following couplings:
\beqa
S &=&-\frac{\gamma}{\kappa^2} \int d^{10}x\sqrt{-G}\bigg[-8 F_{k n,h} F_{m p q s,k} H_{h p r,m} R_{n q r s}+8 F_{m n,h} F_{k p q s,h} H_{m p r,k} R_{n q r s}\nonumber\\&&-2 F_{k n,h} F_{n p q s,k} H_{q r s,m} R_{h r m p}+4 F_{k n,h} F_{m p q s,k} H_{p q r,m} R_{h r n s}-a1 F_{p s,q} F_{h m p s,n} H_{k q r,n} R_{k r h m}\nonumber\\&&-a2 F_{p q,n} F_{h m p s,n} H_{k q r,s} R_{k r h m}+\frac{1}{3} F_{m p,h} F_{n q r s,k} H_{n q s,h} R_{k r m p}-8 F_{n q,h} F_{n p r s,k} H_{h q s,m} R_{m r k p}\nonumber\\&&-4 F_{n q,h} F_{n p r s,k} H_{k p s,m} R_{m r h q}-4 F_{m n,h} F_{k p q s,h} H_{p q r,k} R_{m r n s}-4 F_{n q,h} F_{m p r s,k} H_{p q s,m} R_{n r h k}\nonumber\\&&+4 F_{n q,h} F_{h p r s,k} H_{p q s,m} R_{n r k m}+4 F_{n q,h} F_{m p r s,k} H_{k q s,m} R_{n r h p}-8 F_{n q,h} F_{m p r s,k} H_{h q s,m} R_{n r k p}\nonumber\\&&+F_{m n,h} F_{p q r s,k} H_{k q s,m} R_{p r h n}+2 F_{m p,h} F_{n q r s,k} H_{m q s,k} R_{p r h n}-4 F_{m p,h} F_{n q r s,k} H_{m q s,h} R_{p r k n}\nonumber\\&&+4 F_{m p,h} F_{n q r s,h} H_{m q s,k} R_{p r k n}+4 F_{m p,h} F_{m n q s,k} H_{q r s,h} R_{p r k n}-4 F_{m p,h} F_{m n q s,h} H_{q r s,k} R_{p r k n}\nonumber\\&&-2 F_{n r,h} F_{h n p s,k} H_{p q s,m} R_{q r k m}+4 F_{n q,h} F_{m p r s,k} H_{k p s,m} R_{q r h n}-2 F_{n q,h} F_{m p r s,k} H_{h p s,m} R_{q r k n}\nonumber\\&&+8 F_{m p,h} F_{m n q s,k} H_{p r s,h} R_{q r k n}-8 F_{m p,h} F_{m n q s,h} H_{p r s,k} R_{q r k n}-4 F_{k n,h} F_{n p q s,k} H_{h r s,m} R_{q r m p}\nonumber\\&&-4 F_{n q,h} F_{m p r s,k} H_{h k s,m} R_{q r n p}-4 F_{n r,h} F_{n p r s,k} H_{k p q,m} R_{m s h q}-4 F_{n r,h} F_{n p r s,k} H_{h p q,m} R_{m s k q}\nonumber\\&&+4 F_{n r,h} F_{h n p s,k} H_{p q r,m} R_{m s k q}+4 F_{n r,h} F_{h p r s,k} H_{k p q,m} R_{m s n q}+8 F_{n r,h} F_{h m p s,k} H_{p q r,m} R_{n s k q}\nonumber\\&&+2 F_{m q,h} F_{m n q r,k} H_{p r s,h} R_{p s k n}-4 F_{m q,h} F_{m n q r,h} H_{p r s,k} R_{p s k n}-4 F_{m q,h} F_{k n q r,h} H_{p r s,k} R_{p s m n}\nonumber\\&&-4 F_{n r,h} F_{h n p s,k} H_{p q r,m} R_{q s k m}-4 F_{k n,h} F_{h n p r,k} H_{q r s,m} R_{q s m p}-4 F_{k n,h} F_{h m p r,k} H_{q r s,m} R_{q s n p}\nonumber\\&&-8 F_{n r,h} F_{m n p s,k} H_{k p q,m} R_{r s h q}-8 F_{n r,h} F_{m n p s,k} H_{h p q,m} R_{r s k q}-8 F_{n r,h} F_{h m p s,k} H_{k p q,m} R_{r s n q}\bigg]\nonumber
\eeqa
Using the observation that S-matrix does not produce couplings with zero and one $k_i\inn k_j$, one can find the unknown constants to be  $a1=4$ and $a2=-1$.

\end{document}